\documentclass[twoside,twocolumn,english,pra,twocolumn,aps,superscriptaddress,showpacs]{revtex4-2}

\usepackage[T1]{fontenc}
\usepackage[latin9]{inputenc}
\usepackage{color}
\usepackage{babel}
\usepackage{bm}
\usepackage{amsmath}
\usepackage{amsthm}
\usepackage{graphicx}
\usepackage[pdfusetitle,
 bookmarks=true,bookmarksnumbered=false,bookmarksopen=false,
 breaklinks=false,pdfborder={0 0 0},pdfborderstyle={},backref=false,colorlinks=true]
 {hyperref}

\makeatletter
\theoremstyle{plain}

\usepackage{tabularx}

\usepackage{makecell}
\usepackage{fontenc}

\hypersetup{urlcolor=blue}

\usepackage{cmap}
\usepackage[T1]{fontenc}

\input glyphtounicode
\makeatother

\begin{document}
\title{Toward Heisenberg-Limited Interferometry with Dual Squeezers}
\author{Yi Gu}
\affiliation{Institute of Quantum Information and Technology, Nanjing University
of Posts and Telecommunications, Nanjing 210003, China}
\author{Song-Ping Wang}
\affiliation{Institute of Quantum Information and Technology, Nanjing University
of Posts and Telecommunications, Nanjing 210003, China}
\author{Wei Zhong}
\email{zhongwei1118@gmail.com}

\affiliation{Institute of Quantum Information and Technology, Nanjing University
of Posts and Telecommunications, Nanjing 210003, China}
\begin{abstract}
The canonical Mach-Zehnder interferometer fed with a coherent state
and a squeezed-vacuum state of equal intensities is theoretically
predicted to achieve Heisenberg scaling in phase sensitivity. However,
this ultimate performance is unattainable using direct photon-number-difference
detection due to a divergence arising precisely at the optimal equal-intensity
regime. In this work, we introduce a dual-squeezing approach that
overcomes this fundamental limitation. Our scheme employs an additional
single-mode squeezer before detection, forming a paired configuration
with the input squeezer used to generate the squeezed-vacuum state.
We analytically demonstrate that the resulting dual-squeezing Mach-Zehnder
interferometer enables Heisenberg-limited phase sensitivity with direct
photon-number-difference detection, while remaining robust against
detection noise. Our work provides a feasible and robust route toward
quantum-limited interferometric phase measurements.
\end{abstract}
\maketitle
\emph{Introduction}---Optical two-mode interferometers serve as indispensable
platforms for applications ranging from high-precision metrology \citep{Lee2002JOP,Dowling2008Review,PARIS2009review,OBrien2009Review,Giovannetti2011,DemkowiczDobrzanski2015Book,Barsotti2018RP,Liu2019JPA,Polino2020review,Jin2024review}
to spectroscopy \citep{Leibfried2004SCI,Taylor2016PR,Moreau2019Review,Lee2021CR,Schaffrath2024NJP,Defienne2024review}.
In classical setups, one input port of the interferometer is fed with
a coherent state while the other remains unused resulting in vacuum.
Vacuum fluctuations in the unused port limit the phase sensitivity
to the shot-noise limit (SNL) $1/\sqrt{\bar{n}}$, where $\bar{n}$
is the mean photon number of the coherent state. In 1981, Caves in
a seminal work showed that injecting a squeezed vacuum into the normally
unused port suppresses vacuum fluctuations and enables sub-shot-noise
phase sensitivity \citep{Caves1981PRD}. This breakthrough finding
sparked decades of subsequent technological progress, starting with
early proof-of-principle demonstrations \citep{Xiao1987PRL,Grangier1987PRL},
paving the way for the practical deployment of squeezed light to surpass
the SNL in state-of-the-art gravitational-wave observatories, including
the GEO 600 \citep{LIGO2011} and Hanford LIGO detectors \citep{LIGO2013}.
More recently, it has been rigorously verified that, for fixed total
photon-number resources, squeezed vacuum is the optimal choice for
the unused port when combined with a coherent state in the other input
\citep{Lang2013PRL}. 

The potential of this scheme extends far beyond sub-shot-noise performance.
Remarkably, when the coherent and squeezed-vacuum inputs have equal
intensities, the phase sensitivity approaches the Heisenberg limit
(HL) $1/\bar{n}$, where $\bar{n}$ is the total mean photon number
\citep{Giovannetti2006PRL,Hyllus2010PRL,Giovannetti2011}. However,
this limit is inaccessible with direct photon-number-difference detection
because the signal vanishes at the optimal equal-intensity regime.
Although Bayesian inference can mitigate this issue \citep{Pezze2008PRL,Hofmann2009PRA,Lang2013PRL,Zhong2017PRA},
it requires high-efficiency photon-number-resolving detection \citep{Divochiy2008NP,Sahin2013APL}
and substantial post-processing, thereby imposing severe practical
challenges. Moreover, such approaches are extremely vulnerable to
detection noise \citep{Spagnolo2012PRL,Pezze2013PRL,Oh2017PRA,Gard2017EPJ,Zhong2017PRA},
even small imperfections such as photon loss can significantly degrade
performance, rendering the scheme nearly impractical in realistic
scenarios. 

To overcome these limitations, we propose a modified scheme based
on the original Caves configuration by inserting an additional single-mode
squeezer prior to detection, which can be implemented using an optical
parametric amplifier. The scheme is referred to as a dual-squeezing
Mach-Zehnder interferometer (DS-MZI), as it features a symmetric dual-squeezing
structure, with one squeezer acting at the input side of the interferometer
and the other at the output side. This DS-MZI shares an analogous
dual-squeezing structure with the recently proposed interaction-based
readout implemented in atomic interferometry \citep{Davis2016PRL,Linnemann2016PRL,Nolan2017PRL,Anders2018PRA,Mao2023nphys,Liu2025PRL}.
Beyond the differences between optical and atomic platforms, the two
approaches differ fundamentally in that our scheme involves only local
single-mode squeezing, whereas the previous approach relies on two-internal-mode
squeezing generated by one-axis twisting or two-axis counter-twisting
interactions \citep{Kitagawa1993PRA}.

In this work, we present a complete analytical framework for the DS-MZI,
derive closed-form expressions for the phase sensitivity under photon-number-difference
detection, and show that the scheme removes the divergence at the
optimal operating point while saturating the ultimate phase sensitivity.
We further extend our analysis to detection under imperfect conditions,
such as photon loss, and to unbalanced configurations with different
input and output squeezing strengths. Our results demonstrate that
dual-squeezing protocol provides a simple and experimentally viable
path to noise-resilient Heisenberg-limited interferometry using interferometric
techniques and direct photon-number-difference detection, without
relying on complex exotic state generation \citep{Gerry2000,Boto2000PRL,Krischek2011PRL,Joo2011PRL,Thekkadath2020npj}
or elaborate post-processing \citep{Pezze2008PRL,Zhong2017PRA,Xu2020PRL}.
\begin{figure}[t]
\centering{}\includegraphics[scale=0.47]{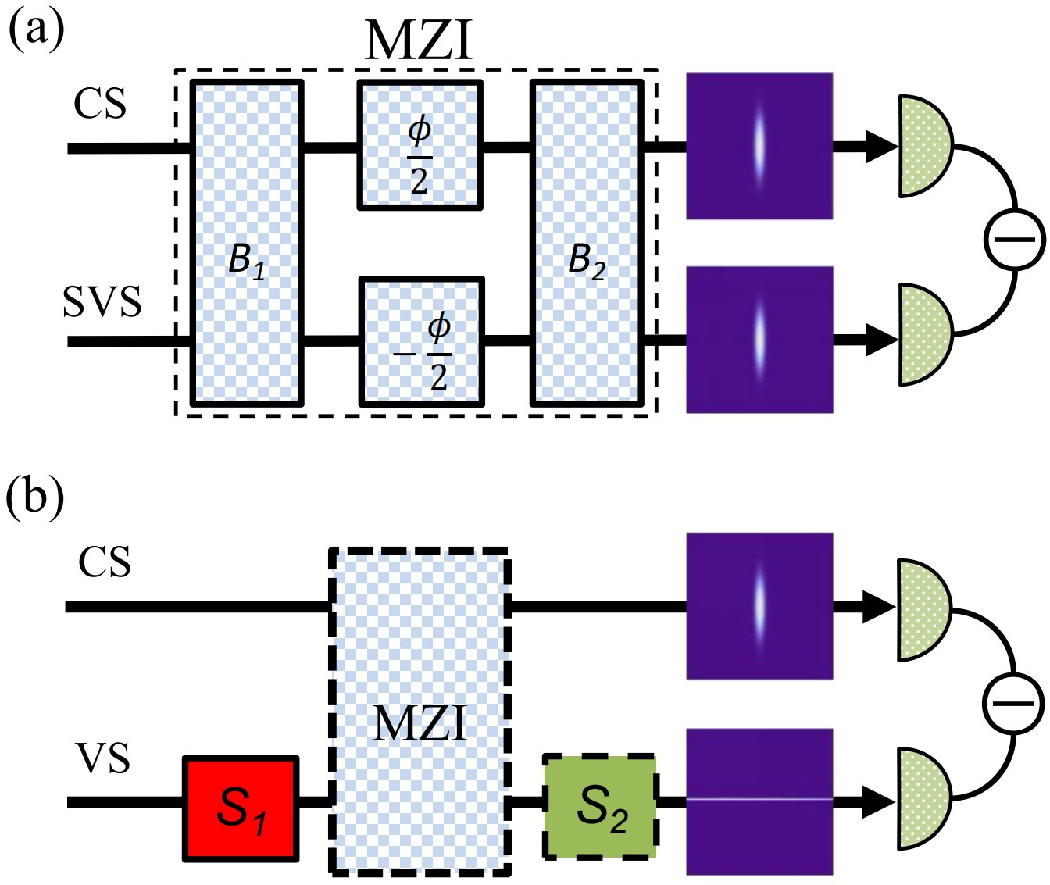}\caption{(Color online) Schematic of (a) the conventional Caves scheme using
a standard Mach-Zehnder interferometer (MZI) fed by a coherent state
(CS) and a squeezed-vacuum state (SVS), and (b) the DS-MZI scheme,
where a vacuum state (VS) is squeezed by $S_{1}$ before entering
the MZI, and an additional squeezer $S_{2}$ is applied to one output
mode prior to detection. The Wigner quasi-probability distributions
for the output ports are plotted for equal input intensities. \label{fig:DS-MZI}}
\end{figure}

\emph{Framework of DS-MZI}---We begin by formalizing the theoretical
framework of the DS-MZI scheme, with the standard MZI as a reference.
The MZI consists of two 50:50 beam splitters $B_{1}$ and $B_{2}$,
and two phase shifters $\phi/2$ and $-\phi/2$ that introduce the
phase difference $\phi$ to be estimated between the two arms. The
total unitary operator of the MZI is $U_{{\rm MZI}}\!=\!B_{2}UB_{1}$.
The DS-MZI is further equipped with two single-mode squeezers $S_{1}$
and $S_{2}$, giving the overall unitary transformation $U_{{\rm DS-MZI}}\!=\!S_{2}U_{{\rm MZI}}S_{1}$,
as illustrated in Fig.~\ref{fig:DS-MZI}.

The two bosonic modes are described by annihilation operators $a$
and $b$, and creation operators $a^{\dag}$ and $b^{\dag}$. Mode
$a$ is injected with a coherent state $\vert\alpha\rangle\!=\!\exp[\alpha(a^{\dag}-a)]\vert0\rangle$
with real-valued amplitude $\alpha$, while mode\textbf{ $b$} is
initially in the vacuum state $\vert0\rangle$ and is then squeezed
by $S_{1}\!=\!\exp[-r(b^{\dag2}-b^{2})/2]$ to produce a squeezed-vacuum
state $\vert r\rangle\!=\!S_{1}\vert0\rangle$, where the squeezing
parameter $r$ is assumed to be real. Here, both $\alpha$ and $r$
are taken to be real to satisfy the optimal phase condition for interferometric
phase estimation (see Appendix A). The output state of the DS-MZI
is $\vert\psi\rangle\!=\!U_{{\rm DS-MZI}}\vert\alpha,0\rangle$. In
the absence of $S_{2}$, this reduces to $U_{{\rm MZI}}S_{1}\vert\alpha,0\rangle\!=\!U_{{\rm MZI}}\vert\alpha,r\rangle$
recovering the original Caves scheme \citep{Caves1981PRD}. 

\emph{Phase sensitivities}---We consider direct photon-number-difference
detection applied at the output ports with the observable $N_{-}\!=\!a^{\dag}a-b^{\dag}b$.
Using the phase-shift operator $U\!=\!\exp[i\phi(a^{\dagger}a-b^{\dagger}b)/2]$
and the 50:50 beam-splitter operator $B_{1}\!=\!\exp[-i\pi(a^{\dagger}b-ab^{\dagger})/4]$,
and imposing the symmetries $B_{1}\!=\!B_{2}$ and $S_{1}\!=\!S_{2}$,
we obtain the expectation value and variance 
\begin{align}
\langle N_{-}\rangle & =\alpha^{2}\Big(\sin^{2}\!\frac{\phi}{2}-\cos^{2}\!\frac{\phi}{2}e^{2r}\Big),\label{eq:expectation}\\
\text{\ensuremath{\langle\Delta^{2}N_{-}}}\rangle & =\alpha^{2}\bigg[\Big(\sin^{2}\!\frac{\phi}{2}-\cos^{2}\!\frac{\phi}{2}e^{2r}\Big)^{2}\!\!+\sin^{2}\!\phi\cosh^{2}\!r\bigg]\quad\nonumber \\
 & \quad+\cos^{4}\!\frac{\phi}{2}\sinh^{2}\!2r,\label{eq:variance}
\end{align}
where $\langle\Delta^{2}N_{-}\rangle\!\equiv\!\langle N_{-}^{2}\rangle-\langle N_{-}\rangle^{2}$.
A detailed derivation is provided in Appendix B. Substituting these
expressions into the error-propagation formula \citep{Yurke1986PRL,Zhong2014JPA}
$\Delta\phi\!=\!\sqrt{\!\langle\Delta^{2}N_{-}\rangle}/\vert d\langle N_{-}\rangle/d\phi\vert$
yields the detection-based phase sensitivity for the DS-MZI 
\begin{figure}[t]
\centering{}\includegraphics[scale=0.9]{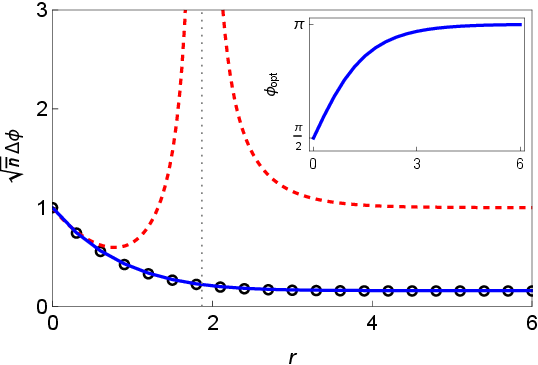}\caption{(Color online) Scaled phase sensitivity $\sqrt{\bar{n}}\Delta\phi$
as a function of the squeezing parameter $r$ for $\alpha=\sqrt{10}$.
The red dashed line represents the phase sensitivity for the Caves
scheme (without $S_{2}$), given by Eq.~(1) in Ref.~\citep{Pezze2008PRL},
while the black circles denote the result from Eq.~\eqref{eq:bound}.
The blue solid line represents the minimum value of Eq.~\eqref{eq:DS-MZI-sensitivity}
with respect to $\phi$, with the inset displaying the corresponding
optimal working points $\phi_{{\rm opt}}$. The vertical gray dotted
line at $r\!\sim\!1.87$ marks the condition $\alpha^{2}\!=\!\sinh^{2}\!r$.
\label{fig:DS-MZI-Sensitivity}}
\end{figure}
\begin{equation}
\Delta\phi=\frac{2g\cosh^{2}\!r}{\alpha(1+e^{2r})},\label{eq:DS-MZI-sensitivity}
\end{equation}
with
\begin{equation}
g=\sqrt{1+\frac{(e^{2r}\cos^{2}\!\frac{\phi}{2}-\sin^{2}\!\frac{\phi}{2})^{2}}{\cosh^{2}\!r\sin^{2}\!\phi}+\frac{\sinh^{2}2r\cos^{4}\frac{\phi}{2}}{\alpha^{2}\cosh^{2}\!r\sin^{2}\!\phi}}.
\end{equation}
Hereafter we omit the factor $1/\!\sqrt{p}$ associated with $p$
independent repetitions to highlight the quantum advantage. The optimal
working point $\phi_{{\rm opt}}$ that minimizes $\Delta\phi$ is
shown numerically in the inset of Fig.~\ref{fig:DS-MZI-Sensitivity}.
In the limit $\alpha^{2}\!\simeq\!\sinh^{2}\!r\!\simeq\!e^{2r}\!/4$,
we find $\phi_{{\rm opt}}\!=\!2\arctan(e^{2r}\!+e^{4r})^{1/4}$. For
comparison, the sensitivity of the Caves scheme (without $S_{2}$)
is obtained from Eq.~(1) in Ref.~\citep{Pezze2008PRL} with $\phi_{{\rm opt}}\!=\!\pi/2$. 

From quantum estimation theory, the ultimate phase sensitivity of
an interferometer fed with coherent-plus-squeezed-vacuum input states
is given by \citep{Pezze2008PRL,Jarzyna2012PRA}, 
\begin{equation}
\Delta\phi=\frac{1}{\sqrt{\!\alpha^{2}e^{2r}+\sinh^{2}\!r}},\label{eq:bound}
\end{equation}
which is independent of $\phi$. The dual-squeezing configuration
uses the same input states and thus retains the same ultimate sensitivity
as the conventional single-squeezing scheme. For a fixed total mean
photon number $\bar{n}\!=\!\alpha^{2}\!+\!\sinh^{2}\!r$, the minimum
is obtained using the method of Lagrange multipliers at $\alpha^{2}\!=\!(e^{2r}\!-\!1)\sinh2r/2e^{2r}$.
Thus, Heisenberg scaling $\Delta\phi\!\sim\!1/\bar{n}$ is achieved
when $\alpha^{2}\!\simeq\!\sinh^{2}\!r\!\simeq\!e^{2r}/4\!\sim\!\bar{n}/2$.

To facilitate quantitative analysis, we introduce two key quantities:
scaled phase sensitivity and saturability. The scaled phase sensitivity
$\sqrt{\bar{n}}\Delta\phi$ is defined as the ratio of the phase sensitivity
to the shot-noise limit, and quantifies the degree of sub-shot-noise
performance for a given metrological protocol. The saturability $S\!\equiv\!(\Delta\phi)_{{\rm bound}}/(\Delta\phi)_{{\rm detection}}$
is defined as the ratio of the ultimate sensitivity bound to the phase
sensitivity achieved by a specific detection, and characterizes how
closely the measurement saturates the fundamental sensitivity limit.
A maximum value $S\!=\!1$ corresponds to an optimal measurement.
For the balanced DS-MZI and moderately large $r$ such that $\sinh^{2}\!r\!\gg\!\alpha^{2}$,
the scaled ultimate phase sensitivity in Eq.~\eqref{eq:bound} approaches
$1/\sqrt{1+4\alpha^{2}}$. The scaled detection-based sensitivity
in Eq.~\eqref{eq:DS-MZI-sensitivity} attains a minimum value $1/(2\alpha)$
with $g\!\rightarrow\!1$ as $\phi\!\rightarrow\!\pi$. Consequently,
the saturability $S$ is independent of $r$ and increases with $\alpha$,
exceeding $99\%$ for $\alpha\!=\!4$. 

We plot in Fig.~\ref{fig:DS-MZI-Sensitivity} the scaled phase sensitivity
$\sqrt{\bar{n}}\Delta\phi$ as a function of $r$ for fixed $\alpha\!=\!\sqrt{10}$.
Photon-number-difference detection without $S_{2}$ achieves the ultimate
sensitivity bound for $r\!\ll\!1$ corresponding to $\sinh^{2}\!r\!\ll\!\alpha^{2}$
\citep{Ataman2018PRA}, but diverges sharply near $r\!\sim\!1.87$
where $\alpha^{2}\!=\!\sinh^{2}\!r$ \citep{Pezze2008PRL}. This divergence
arises because the expectation value $\langle N_{-}\rangle\!=\!(\alpha^{2}-\sinh^{2}\!r)\cos\phi$
vanishes in the optimal regime, rendering the signal insensitive to
$\phi$. In contrast, the dual-squeezing configuration yields a nonzero
$\langle N_{-}\rangle$ (see Eq.~\eqref{eq:expectation}). This critical
distinction can be visualized in phase space, as shown in Fig.~\eqref{fig:DS-MZI}.
In the conventional configuration (without $S_{2}$), the Wigner quasi-probability
distribution $W(x_{i},p_{i})$ is identical for the two output modes
$i=a,b$, resulting in $\langle N_{-}\rangle\!=\!0$. In contrast,
the presence of $S_{2}$ produces distinct Wigner distributions for
the two output modes and thus a non-vanishing intensity difference
(see Appendix C for details). The dual-squeezing configuration eliminates
the divergence in sensitivity and enables a near-Heisenberg scaling.
For $\alpha\!=\!\sqrt{10}$, the saturability $S$ reaches about $98\%$
at $r\!\sim\!1.87$ and remains at this level as $r$ increases.

\emph{Imperfect detection}---Next, we extend the discussion to imperfect
detection arising from photon loss. Such effects are quantified by
a non-unit detection efficiency $\eta$ \citep{DallArno2010PRA,Lee2021CR},
meaning only a fraction $\eta$ of incoming photons generate an electrical
signal and thus a detection count, while the remainder are lost or
unaccounted for. Here we assume that the detectors acting on each
output port have identical detection efficiency. These imperfections
increase the total noise on the measured signal, modifying the error-propagation
formula under photon-number-difference detection as detailed in Appendix
C 
\begin{figure}[t]
\begin{centering}
\includegraphics[scale=0.9]{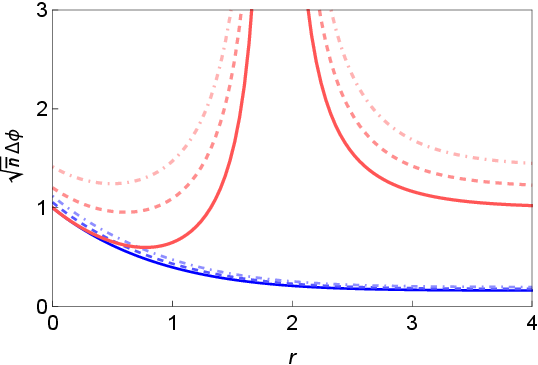}
\par\end{centering}
\centering{}\caption{(Color online) Scaled phase sensitivity $\sqrt{\bar{n}}\Delta\phi$
as a function of the squeezing parameter $r$ under imperfect detection
for $\alpha\!=\!\sqrt{10}$. Red lines correspond to the conventional
MZI scheme (without $S_{2}$) while blue lines represent the DS-MZI
scheme. Within each color set, curves from light to dark indicate
detection efficiencies $\eta\!=\!0.8$, $0.9$, and the ideal case
$\eta\!=\!1$, respectively. \label{fig:NoisyDetection}}
\end{figure}
\begin{equation}
\Delta\phi=\frac{\sqrt{\langle\Delta^{2}N_{-}\rangle+\frac{1-\eta}{\eta}\langle N_{+}\rangle}}{\left|\frac{d\langle N_{-}\rangle}{d\phi}\right|},\label{eq:error}
\end{equation}
where $N_{+}\!=\!a^{\dag}a+b^{\dag}b$ denotes the total output photon-number
operator. In the numerator, the first term $\langle\Delta^{2}N_{-}\rangle$
accounts for quantum projection noise, while the second term $(1-\eta)\langle N_{+}\rangle/\eta$
represents the contribution of detection noise.

For the conventional MZI scheme (without $S_{2}$), all optical elements---including
beam splitters and phase shifters---are passive, so the total mean
photon number is conserved, i.e., $\langle N_{+}\rangle\!=\!\bar{n}\!=\!\alpha^{2}+\sinh^{2}\!r$.
This implies that the detection noise depends solely on $\eta$ for
fixed $\bar{n}$, which is consistent with the technical noise considered
in prior works \citep{Davis2016PRL,Szigeti2017PRL}. In contrast,
in the DS-MZI the second squeezer acts as an active optical element
that modifies the total output photon number, such that $\langle N_{+}\rangle\!\neq\!\bar{n}$.
The explicit expression for $\langle N_{+}\rangle$ in the DS-MZI
(see Appendix B for details) is 
\begin{align}
\langle N_{+}\rangle & =\alpha^{2}\Big(\!\sin^{2}\!\frac{\phi}{2}+\cos^{2}\!\frac{\phi}{2}e^{2r}\!\Big)+\cos^{2}\!\frac{\phi}{2}(\cosh2r-1).\label{eq:TotalPhotonNumber}
\end{align}

As illustrated in Fig.~\ref{fig:NoisyDetection}, the scaled phase
sensitivity $\sqrt{\bar{n}}\Delta\phi$ under imperfect detection
is plotted as a function of $r$ for $\alpha\!=\!\sqrt{10}$. The
figure clearly shows that the conventional MZI scheme is highly vulnerable
to detection noise. For non-ideal efficiencies $\eta\!=\!0.8$ and
$0.9$, the scaled sensitivity degrades significantly compared to
the ideal case $\eta\!=\!1$, and the scheme loses its sub-shot-noise
scaling capability. Notably, the optimal working point for the conventional
scheme remains fixed at $\phi_{{\rm opt}}\!=\!\pi/2$ for all values
of $\eta$. In sharp contrast, the dual-squeezing MZI configuration
is remarkably robust against detection noise, with its sensitivity
exhibiting only weak dependence on $\eta$. The curves for $\eta\!=\!0.8$
and $0.9$ almost overlap with the ideal case $\eta\!=\!1$, particularly
at larger squeezing strength $r$. This behavior can be explicitly
understood from Eq.~\eqref{eq:error}, that for large $r$, the detection-noise
contribution $(1-\eta)\langle N_{+}\rangle/\eta$ becomes negligible
compared to the quantum noise term $\langle\Delta^{2}N_{-}\rangle$.
As a result, the detection sensitivity given by Eq.~\eqref{eq:error}
becomes effectively independent of $\eta$. Similarly, numerical analysis
confirms that the optimal working point for the DS-MZI under imperfect
detection remains consistent with the ideal case $\eta\!=\!1$, as
shown in the inset of Fig.~\ref{fig:DS-MZI-Sensitivity}. These results
highlight the inherent resilience of the proposed dual-squeezing scheme.
It is worth emphasizing that, while optical amplification (as applied
in our scheme $S_{2}$) has been normally employed to alleviate the
impact of imperfect detectors \citep{DallArno2010PRA}, our scheme
realizes efficient noise suppression with only a single optical amplifier
$S_{2}$, rather than introducing one amplifier for each noisy detector. 

\emph{Enhancing sensitivity via unbalanced configuration}---In the
previous sections, we assumed the symmetric condition $S_{1}\!=\!S_{2}$
for simplicity. Here we relax this constraint and consider the more
general case of an unbalanced DS-MZI ($S_{1}\!\neq\!S_{2}$) to identify
the optimal configuration for achieving the best phase sensitivity
by trading off the input and output squeezing parameters $r_{1}$
and $r_{2}$. For our analysis, we rederive the expectation value
and variance of the intensity-difference observable $N_{-}$, as well
as the total photon-number operator $N_{+}$, for the unbalanced case.
Using the unitary operator $U_{{\rm DS-MZI}}\!=\!S_{2}U_{{\rm MZI}}S_{1}$,
where $S_{1}\!=\!e^{-r_{1}(b^{\dag2}-b^{2})/2}$ and $S_{2}\!=\!e^{-r_{2}(b^{\dag2}-b^{2})/2}$,
and also imposing $B_{1}\!=\!B_{2}$, we obtain (see Appendix B for
details)
\begin{align}
\langle N_{-}\rangle & =\alpha^{2}\Big(\!\sin^{2}\!\frac{\phi}{2}-\cos^{2}\!\frac{\phi}{2}e^{2r_{2}}\!\Big)-\sin^{2}\!\frac{\phi}{2}\sinh^{2}\!(r_{1}-r_{2})\nonumber \\
 & \quad+\frac{1}{2}\cos^{2}\!\frac{\phi}{2}(\cosh\!2r_{1}-\cosh\!2r_{2}),\label{eq:expNmins}
\end{align}
and
\begin{align}
\langle N_{+}\rangle & =\alpha^{2}\Big(\!\sin^{2}\!\frac{\phi}{2}+\cos^{2}\!\frac{\phi}{2}e^{2r_{2}}\!\Big)+\sin^{2}\!\frac{\phi}{2}\sinh^{2}\!(r_{1}-r_{2})\nonumber \\
 & \quad+\frac{1}{2}\cos^{2}\!\frac{\phi}{2}(\cosh2r_{1}+\cosh2r_{2}-2),\label{eq:expNplus}
\end{align}
\begin{widetext}
\begin{align}
\text{\ensuremath{\langle\Delta^{2}N_{-}}}\rangle & =\alpha^{2}\bigg[\Big(e^{2\text{\ensuremath{r_{2}}}}\cos^{2}\!\frac{\phi}{2}-\sin^{2}\!\frac{\phi}{2}\Big)^{2}+\frac{1}{4}(e^{2\text{\ensuremath{r_{2}}}-r_{1}}+e^{-\text{\ensuremath{r_{1}}}})^{2}\sin^{2}\!\phi\bigg]+\Big[\frac{1}{2}\sinh^{2}\!2r_{1}+2\sinh^{2}(2r_{1}-r_{2})\cosh^{2}\!r_{2}\Big]\cos^{4}\frac{\phi}{2}\nonumber \\
 & \quad+\Big[\cosh^{2}\text{\ensuremath{r_{2}}}-\cosh(4\text{\ensuremath{r_{1}}}-3\text{\ensuremath{r_{2}}})\cosh\text{\ensuremath{r_{2}}}\Big]\cos^{2}\!\frac{\phi}{2}+\sinh^{2}(\text{\ensuremath{r_{1}}}-\text{\ensuremath{r_{2}}})\cosh^{2}\!\text{\ensuremath{r_{2}}}\sin^{2}\!\phi+\frac{1}{4}\big[\cosh(4\text{\ensuremath{r_{1}}}-4\text{\ensuremath{r_{2}}})-1\big],\label{eq:varNmins}
\end{align}
\begin{figure}[t]
\begin{centering}
\includegraphics[scale=0.67]{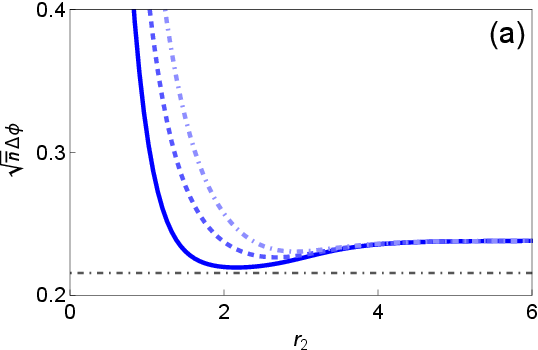}\includegraphics[scale=0.64]{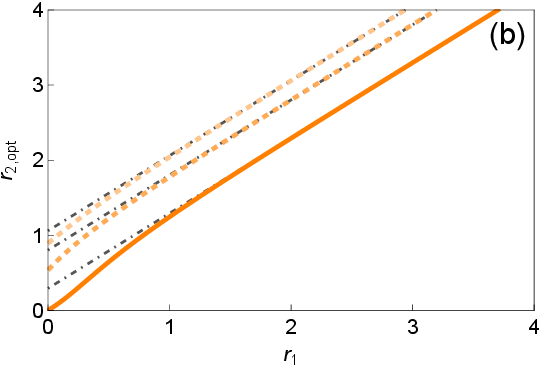}\includegraphics[scale=0.64]{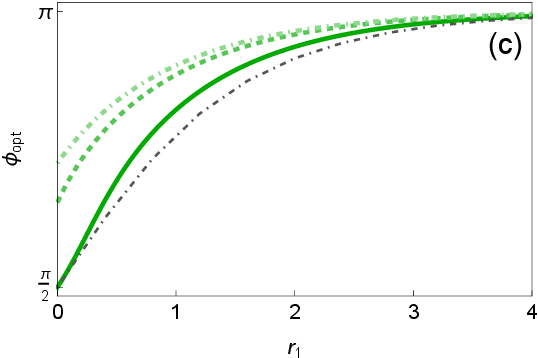}
\par\end{centering}
\centering{}\caption{(Color online) Performance comparison of the unbalanced DS-MZI under
imperfect detection for $\alpha\!=\!\sqrt{10}$. In panel (a), the
input squeezing parameter $r_{1}$ is fixed at $r_{1}\!=\!1.87$,
which lies in the equal-intensity regime $\alpha^{2}\!=\!\sinh^{2}\!r_{1}$.
From light to dark, the lines represent detection efficiency $\eta\!=\!0.8$,
$0.9$ and the ideal case $\eta\!=\!1$, respectively. (a) Scaled
phase sensitivity $\sqrt{\bar{n}}\Delta\phi$ as a function of the
output squeezing parameter $r_{2}$. The horizontal gray dot-dashed
line denotes the ultimate sensitivity bound given by Eq.~\eqref{eq:bound}.
(b) Optimal output squeezing parameter $r_{2,{\rm opt}}$ as a function
of the input squeezing parameter $r_{1}$, with $\phi$ optimized
simultaneously. The gray dot-dashed lines are fits to $r_{1}\!+\!\delta$,
with offsets $\delta\!\sim\!0.29$, $0.80$ and $1.06$ (from darker
to lighter, corresponding to $\eta\!=\!1$, $0.9$, and $0.8$, respectively).
(c) Optimal working point $\phi_{{\rm opt}}$ as a function of the
input squeezing parameter $r_{1}$, with $r_{2}$ optimized simultaneously.
The gray dot-dashed line refers to the balanced case $r_{1}\!=\!r_{2}$,
corresponding to the result shown in the inset of Fig.~\ref{fig:DS-MZI-Sensitivity}.
\label{fig:Tradeoff}}
\end{figure}
\end{widetext}Substituting these expressions into the standard error-propagation
formula yields the detection-based sensitivity for the unbalanced
DS-MZI scheme, although the resulting expression is too cumbersome
to present explicitly. Notably, the output squeezing parameter $r_{2}$
plays a dominant role in determining the sensitivity in the unbalanced
case, in contrast to the balanced scenario. When $r_{2}\!=\!r_{1}=\!r$,
the expressions reduce to those for the balanced DS-MZI scheme. When
$r_{2}\!=\!0$, they recover the phase sensitivity for the conventional
MZI scheme (without $S_{2}$), as given by Eq.~(1) in Ref.~\citep{Pezze2008PRL}.
Substituting these into the modified error-propagation formula in
Eq.~\eqref{eq:error} extends our analysis to scenarios with imperfect
detection.

As illustrated in Fig.~\ref{fig:Tradeoff} (a), we plot the scaled
phase sensitivity $\sqrt{\bar{n}}\Delta\phi$ as a function of $r_{2}$
for $\alpha\!=\!\sqrt{10}$ and fixed $r_{1}\!=\!1.87$. An unbalanced
DS-MZI configuration can further enhance the phase sensitivity relative
to the balanced case. For the ideal case ($\eta\!=\!1$), the scaled
phase sensitivity decreases with increasing $r_{2}$, reaches a minimum
(with saturability $S\!>\!98\%$), and then rises to a stable plateau
(saturability $S\!\sim\!91\%$). For the non-ideal detection efficiencies
($\eta\!=\!0.9,0.8$), the saturability at the minima is approximately
$S\sim95\%$ and $94\%$, respectively. The minima shift towards larger
$r_{2}$ and gradually disappear as $\eta$ decreases, after that,
the sensitivity decreases monotonically before settling at the same
plateau level observed for $\eta\!=\!1$. The plateau is approximately
governed by $5\sqrt{19+\cosh\!2r_{1}}/(20e^{r_{1}}+\sinh\!r_{1})$
independent of $\eta$ at the optimal phase point $\phi_{{\rm opt}}\!=\!2\arctan\!3e^{r_{1}}$,
which confirms that an unbalanced DS-MZI with $r_{2}\!>\!r_{1}$ possesses
inherent resilience to detection noise. A sharp divergence occurs
at $r_{2}\!=\!0$, as the DS-MZI with vanishing $r_{2}$ reduces to
the conventional MZI scheme, which exhibits divergence for the parameters
$\alpha\!=\!\sqrt{10}$ and $r_{1}\!=\!1.87$, corresponding to the
equal input-intensity regime. This constitutes the primary motivation
of the present work.

Figures~\ref{fig:Tradeoff}(b) and (c) show the optimal output squeezing
parameter $r_{2,{\rm opt}}$ and the optimal working point $\phi_{{\rm opt}}$
as functions of the input squeezing parameter $r_{1}$. From Fig.~\ref{fig:Tradeoff}(b),
it is revealed that a larger $r_{2}$ relative to $r_{1}$ (i.e.,
$r_{2}\!>\!r_{1}$) is an optimal choice under both ideal and non-ideal
conditions, indicating that an unbalanced configuration yields the
best sensitivity. We define the relative offset by $\delta\!\equiv\!r_{2,{\rm opt}}\!-r_{1}$.
For the ideal case ($\eta\!=\!1$), the offset is $\delta\!\sim\!0.29$
when $r_{1}\!>\!1$. For non-ideal cases, the offsets increase correspondingly
to $\delta\!\sim\!0.80$ and $1.06$, for $\eta\!=\!0.9$ and $0.8$,
respectively. When $r_{1}\!<\!1$, the behavior is different. Specifically,
in the limit $r_{1}\!\rightarrow\!0$ (corresponding to $\alpha^{2}\!\gg\!\sinh^{2}r_{1}$),
for the noiseless case $\eta\!=\!1$, a balanced configuration (i.e.,
$r_{2,{\rm opt}}\!=\!r_{1}$) is optimal, while for the noisy case
$\eta\!\neq\!1$, unbalanced configurations are also optimal with
offsets $\delta\!=\!0.54$ and $0.89$ corresponding to $\eta\!=\!0.9$
and $0.8$, respectively. As shown in Fig.~\ref{fig:Tradeoff}(c),
under noiseless conditions, the optimal working point $\phi_{{\rm opt}}$
for both unbalanced and balanced configurations increases monotonically
from $\pi/2$ to $\pi$ as $r_{1}$ increases, with the unbalanced
case exhibiting a relatively larger curvature than the balanced one.
Under noisy conditions, $\phi_{{\rm opt}}$ remains monotonic but
no longer starts at $\pi/2$.

\emph{Conclusion}---We propose a dual-squeezing Mach-Zehnder interferometer
that eliminates the vanishing intensity-difference signal that plagues
the conventional Caves scheme. Our approach modifies the standard
setup by simply inserting an additional squeezer (optical parametric
amplifier) at the output port before detection. The scheme offers
two key advantages: it enables Heisenberg-limited phase sensitivity
using direct photon-number-difference detection, and it is highly
robust against realistic detection imperfections, with high saturability
of the ultimate sensitivity preserved even under strong detection
noise. Without requiring high-efficiency detectors or complicated
post-processing, the proposed scheme is compatible with standard optical
interferometric setups and readily implementable with existing technology.
Our work has significant implications for practical optical quantum-enhanced
metrology.

\section*{Acknowledgments}

This work was supported by the National Natural Science Foundation
of China through Grant No. 12005106.

\bibliographystyle{apsrev4-1}
\bibliography{ZW}

\onecolumngrid

\section*{Appendix A: Optimal phase for input states \label{sec:AppendixA}}

  \makeatletter \renewcommand{\theequation}{A\arabic{equation}} \makeatother \setcounter{equation}{0}For
a two-mode optical interferometer, the general beam-splitter operator
reads
\begin{equation}
B=e^{-i(\tau a^{\dagger}b+\tau^{\ast}ab^{\dagger})},
\end{equation}
where $\tau\!=\!\left|\tau\right|e^{i\varphi}$ denotes the complex
transmittance \citep{Gerry2004Book}. We focus on balanced 50:50 beam
splitters, for which $\left|\tau\right|\!=\!\pi/4$ , so that
\begin{equation}
B=e^{-i\frac{\pi}{4}(e^{i\varphi}a^{\dagger}b+e^{-i\varphi}ab^{\dagger})}.\label{eq:BS-1}
\end{equation}
A general coherent state is $\vert\alpha\rangle\!=\!D(\alpha)\vert0\rangle$,
with displacement operator 
\begin{equation}
D(\alpha)=e^{\alpha a^{\dag}-\alpha^{\ast}a},
\end{equation}
and complex amplitude $\alpha\!=\!\left|\alpha\right|e^{i\theta_{a}}$.
A general squeezed-vacuum state is $\vert\xi\rangle\!=\!S(\xi)\vert0\rangle$
with squeezing operator 
\begin{equation}
S(\xi)=e^{(\xi b^{\dag2}-\xi^{\ast}b^{2})/2},
\end{equation}
 and complex squeezing parameter $\xi\!=\!re^{i\theta_{b}}$. 

According to quantum estimation theory, the phase sensitivity for
estimating $\phi$ is lower bounded by the quantum Cram\'er-Rao bound
\begin{equation}
\Delta\phi\geq\frac{1}{\sqrt{pF_{Q}}},
\end{equation}
where $F_{Q}$ is the quantum Fisher information and $p$ is the number
of independent experiments \citep{Helstrom1976Book,Holevo1982Book}.
Maximizing $F_{Q}$ yields the best phase sensitivity. In a lossless
interferometer, the quantum Fisher information can be expressed as
\begin{equation}
F_{Q}=4(\langle J_{z}^{2}\rangle-\langle J_{z}\rangle^{2}),
\end{equation}
where $J_{z}\!\equiv\!(a^{\dagger}a-b^{\dagger}b)/2$ is the generator
of the phase-shift operator $U\!=\!e^{i\phi J_{z}}$. For an interferometer
fed with a coherent state and a squeezed-vacuum state $\vert\alpha,\xi\rangle$,
the quantum Fisher information is maximized and thus the sensitivity
achieves the bound given by Eq.~\eqref{eq:bound} when the optimal
phase condition
\begin{equation}
2\theta_{a}-\theta_{b}=\pm2\varphi,
\end{equation}
is satisfied \citep{Liu2013PRA,Ataman2019PRA,Zhong2020SC}. In the
main text, we restrict to the choice $\theta_{a}\!=\!0$, $\theta_{b}\!=\!\pi$,
and $\varphi\!=\!\pi/2$, which fulfills this optimal-phase requirement.

\section*{Appendix B: Mode operators of the DS-MZI in the Heisenberg picture\label{sec:AppendixB}}

  \makeatletter \renewcommand{\theequation}{B\arabic{equation}}\makeatother \setcounter{equation}{0} 

To facilitate our calculations, we first introduce the two-mode SU(2)
Schwinger operators \citep{Ban1993JOSAB}
\begin{equation}
J_{x}\!=\!\frac{a^{\dagger}b+ab^{\dagger}}{2},J_{y}\!=\!\frac{a^{\dagger}b-ab^{\dagger}}{2i},J_{z}=\frac{a^{\dagger}a-b^{\dagger}b}{2},\quad
\end{equation}
the two-mode SU(1,1) operators 
\begin{equation}
K_{x}\!=\!\frac{a^{\dagger}b^{\dagger}+ab}{2},K_{y}\!=\!\frac{a^{\dagger}b^{\dagger}-ab}{2i},K_{z}=\frac{a^{\dagger}a+bb^{\dagger}}{2}.\quad\quad
\end{equation}
and the single-mode SU(1,1) operators for mode $a$
\begin{equation}
K_{a,x}\!=\!\frac{a^{\dagger}a^{\dagger}+aa}{4},K_{a,y}\!=\!\frac{a^{\dagger}a^{\dagger}-aa}{4i},K_{a,z}=\frac{a^{\dagger}a+aa^{\dagger}}{4},
\end{equation}
with analogous definitions for mode $b$. All these operators are
Hermitian, guaranteeing real expectation values and greatly simplifying
the algebra compared to direct products of creation and annihilation
operators.

For the DS-MZI, the optical modes evolve through successive components
in the order
\begin{equation}
a_{0},b_{0}\overset{S_{1}}{\rightarrow}a_{1},b_{1}\overset{B_{1}}{\rightarrow}a_{2},b_{2}\overset{U}{\rightarrow}a_{3},b_{3}\overset{B_{2}}{\rightarrow}a_{4},b_{4}\overset{S_{2}}{\rightarrow}a_{5},b_{5},\qquad
\end{equation}
where the subscripts label the modes at each stage for clarity. The
input--output relations for each element are
\begin{align}
a_{5} & =a_{4},\quad b_{5}=\cosh\!r_{2}b_{4}+\sinh\!r_{2}b_{4}^{\dagger},\\
a_{4} & =\frac{1}{\sqrt{2}}(a_{3}+b_{3}),\quad b_{4}=-\frac{1}{\sqrt{2}}(a_{3}-b_{3}),\\
a_{3} & =a_{2}e^{i\phi/2},\quad b_{3}=b_{2}e^{-i\phi/2},\\
a_{2} & =\frac{1}{\sqrt{2}}(a_{1}+b_{1}),\quad b_{2}=-\frac{1}{\sqrt{2}}(a_{1}-b_{1}),\\
a_{1} & =a_{0},\quad b_{1}=\cosh\!r_{1}b_{0}+\sinh\!r_{1}b_{0}^{\dagger}.
\end{align}
Here, we have set $B_{1}\!=\!B_{2}$ throughout the main text. 

Assuming equal squeezing strengths $r_{1}\!=\!r_{2}\!=\!r$, the output
operators in the Heisenberg picture simplify to 
\begin{align}
a_{5} & =i\sin\!\frac{\phi}{2}a_{0}+\cos\!\frac{\phi}{2}(\cosh\!rb_{0}+\sinh\!rb_{0}^{\dagger}),\\
b_{5} & =-i\sin\!\frac{\phi}{2}b_{0}-\cos\!\frac{\phi}{2}(\cosh\!ra_{0}+\sinh\!ra_{0}^{\dagger}).
\end{align}
We then compute the photon-number operators $a_{5}^{\dagger}a_{5}$
and $b_{5}^{\dagger}b_{5}$. Using the SU(2) and SU(1,1) operators
defined above, the photon-number-difference operator $N_{-}$ and
the total photon-number operator $N_{+}$ take the compact form
\begin{align*}
N_{-} & =a_{5}^{\dagger}a_{5}-b_{5}^{\dagger}b_{5}=h_{-}J_{z}+h_{1}J_{y}+h_{3}(K_{b,x}-K_{a,x}),
\end{align*}
and
\begin{align}
N_{+} & =a_{5}^{\dagger}a_{5}+b_{5}^{\dagger}b_{5}=h_{+}K_{z}+h_{2}K_{y}+h_{3}(K_{b,x}+K_{a,x})-1,
\end{align}
where
\begin{align}
h_{\pm} & =2\big(\sin^{2}\!\frac{\phi}{2}\pm\cos^{2}\!\frac{\phi}{2}\cosh\!2r\big),\\
h_{1} & =2\sin\!\phi\cosh\!r,\\
h_{2} & =2\sin\!\phi\sinh\!r,\\
h_{3} & =2\cos^{2}\!\frac{\phi}{2}\sinh\!2r.
\end{align}
Thus Eqs.~\eqref{eq:expectation}, \eqref{eq:variance} and \eqref{eq:TotalPhotonNumber}
in the main text are derived from the above results. 

For the general case $r_{1}\!\neq\!r_{2}$, the output operators become
\begin{align}
a_{5} & =i\sin\!\frac{\phi}{2}a_{0}+\cos\!\frac{\phi}{2}(\cosh\!r_{1}b_{0}+\sinh\!r_{1}b_{0}^{\dagger}),\\
b_{5} & =-i\sin\!\frac{\phi}{2}\big[\cosh(r_{1}-r_{2})b_{0}+\sinh(r_{1}-r_{2})b_{0}^{\dagger}\big]-\cos\!\frac{\phi}{2}(\cosh\!r_{2}a_{0}+\sinh\!r_{2}a_{0}^{\dagger}).
\end{align}
Accordingly, the photon-number operators $N_{-}$ and $N_{+}$ read
\begin{align}
N_{-} & =k_{1}^{+}K_{z}+k_{2}^{-}J_{z}+k_{3}^{-}K_{a,x}+k_{4}^{-}K_{b,x}+k_{5}^{+}J_{y}+k_{6}^{-}K_{y},
\end{align}
and
\begin{align}
N_{+} & =k_{2}^{+}K_{z}+k_{1}^{-}J_{z}+k_{3}^{+}K_{a,x}+k_{4}^{+}K_{b,x}+k_{5}^{-}J_{y}+k_{6}^{+}K_{y}-1,
\end{align}
where
\begin{align}
k_{1}^{\pm} & =-2\sin^{2}\!\frac{\phi}{2}\sinh^{2}(r_{1}-r_{2})\pm\cos^{2}\!\frac{\phi}{2}(\cosh2r_{1}-\cosh2r_{2}),\\
k_{2}^{\pm} & =2\sin^{2}\!\frac{\phi}{2}\cosh^{2}(r_{1}-r_{2})\pm\cos^{2}\!\frac{\phi}{2}(\cosh2r_{1}+\cosh2r_{2}),\\
k_{3}^{\pm} & =\pm2\cos^{2}\!\frac{\phi}{2}\sinh\!2r_{2},\\
k_{4}^{\pm} & =2\Big[\cos^{2}\!\frac{\phi}{2}\sinh\!2r_{1}\pm\sin^{2}\!\frac{\phi}{2}\sinh2(r_{1}-r_{2})\Big],\\
k_{5}^{\pm} & =\sin\!\phi\big[\cosh\!r_{1}\pm\cosh(2r_{2}-r_{1})\big],\\
k_{6}^{\pm} & =\sin\!\phi\big[\sinh r_{1}\pm\sinh(2r_{2}-r_{1})\big].
\end{align}
Hence, Eqs.~\eqref{eq:expNmins}, \eqref{eq:expNplus} and \eqref{eq:varNmins}
in the main text are derived from the above results. 

\section*{Appendix C: DS-MZI in phase space \label{sec:AppendixC}}

  \makeatletter \renewcommand{\theequation}{C\arabic{equation}}\makeatother \setcounter{equation}{0} In
this appendix, we derive the Wigner function for the DS-MZI at the
output mode. In general, for an arbitrary $n$-mode optical system,
the quadrature operators for mode $k$ are defined as 
\begin{equation}
\hat{x}_{k}=\frac{1}{\sqrt{2}}(a_{k}+a_{k}^{\dag}),\quad\hat{p}_{k}=\frac{1}{i\sqrt{2}}(a_{k}-a_{k}^{\dag}).
\end{equation}
They define the operator vector $\hat{\bm{r}}\!=\!(\hat{x}_{1},\hat{p}_{1},\hat{x}_{2},\hat{p}_{2},\cdots)^{{\rm T}}$.
For Gaussian states of arbitrary $n$ modes, the Wigner function with
the phase-space vector $\bm{r}\!=\!(x_{1},p_{1},x_{2},p_{2},\cdots)^{{\rm T}}$
reads
\begin{equation}
W(\bm{r})=\frac{1}{(2\pi)^{n}\sqrt{\det\sigma}}e^{-\frac{1}{2}(\bm{r}-\bar{\bm{r}})^{{\rm T}}\sigma^{-1}(\bm{r}-\bar{\bm{r}})}.
\end{equation}
The Wigner function is determined by the covariance matrix $\sigma$
and the mean vector $\bar{\bm{r}}$, of which the elements are defined
by 
\begin{equation}
\sigma_{kl}=\frac{1}{2}\langle[\hat{\bm{r}}_{k},\hat{\bm{r}}_{l}]_{+}\rangle-\langle\hat{\bm{r}}_{k}\rangle\langle\hat{\bm{r}}_{l}\rangle,
\end{equation}
where $[\bullet,\bullet]_{+}$ denotes the anti-commutator, and the
mean vector is $\bar{\bm{r}}\!=\!\langle\hat{\bm{r}}\rangle$. In
phase space, any Gaussian unitary evolution in Hilbert space corresponds
to a symplectic transformation described by the symplectic matrix
$F$ and the displacement vector $\bm{d}$
\begin{equation}
\sigma\rightarrow F\sigma F^{{\rm T}},\quad\bar{\bm{r}}\rightarrow F\bar{\bm{r}}+\bm{d}.
\end{equation}

For the DS-MZI, the mode number is $n\!=\!2$ and the two modes correspond
to output modes $a$ and $b$. All optical elements preserve the Gaussian
statistics with zero displacement, i.e., $\bm{d}\!=\!0$. The symplectic
transformation corresponding to the squeezer $S$, 50:50 beam splitter
$B$, and the phase shifter $U$ in the main text are \citep{Gard2017EPJ}
\begin{equation}
F_{S}=\left(\begin{array}{cccc}
1 & 0 & 0 & 0\\
0 & 1 & 0 & 0\\
0 & 0 & e^{r} & 0\\
0 & 0 & 0 & e^{-r}
\end{array}\right),\quad F_{B}=\frac{1}{\sqrt{2}}\left(\begin{array}{cccc}
1 & 0 & 1 & 0\\
0 & 1 & 0 & 1\\
-1 & 0 & 1 & 0\\
0 & -1 & 0 & 1
\end{array}\right),
\end{equation}
and 
\begin{equation}
F_{U}=\left(\begin{array}{cccc}
\cos\frac{\phi}{2} & -\sin\frac{\phi}{2} & 0 & 0\\
\sin\frac{\phi}{2} & \cos\frac{\phi}{2} & 0 & 0\\
0 & 0 & \cos\frac{\phi}{2} & \sin\frac{\phi}{2}\\
0 & 0 & -\sin\frac{\phi}{2} & \cos\frac{\phi}{2}
\end{array}\right).
\end{equation}
For the full DS-MZI sequence, the transformation chain reads 
\begin{equation}
\sigma_{0},\bar{\bm{r}}_{0}\overset{F_{S}}{\rightarrow}\sigma_{1},\bar{\bm{r}}_{1}\overset{F_{B}}{\rightarrow}\sigma_{2},\bar{\bm{r}}_{2}\overset{F_{U}}{\rightarrow}\sigma_{3},\bar{\bm{r}}_{3}\overset{F_{B}}{\rightarrow}\sigma_{4},\bar{\bm{r}}_{4}\overset{F_{S}}{\rightarrow}\sigma_{5},\bar{\bm{r}}_{5}.\qquad
\end{equation}
For initial state $\vert\alpha,0\rangle$, the corresponding covariance
matrix and displacement vector are
\begin{equation}
\sigma_{0}=\left(\begin{array}{cccc}
1 & 0 & 0 & 0\\
0 & 1 & 0 & 0\\
0 & 0 & 1 & 0\\
0 & 0 & 0 & 1
\end{array}\right),\bar{\bm{r}}_{0}=\left(\begin{array}{c}
\sqrt{2}\alpha\\
0\\
0\\
0
\end{array}\right).
\end{equation}
We obtain the covariance matrix and displacement vector for the output
state as
\begin{equation}
\sigma_{5}=F_{S}F_{B}F_{U}F_{B}F_{S}\sigma_{0}F_{S}^{{\rm T}}F_{B}^{{\rm T}}F_{U}^{{\rm T}}F_{B}^{{\rm T}}F_{S}^{{\rm T}},
\end{equation}
\begin{equation}
\bar{\bm{r}}_{5}=F_{S}F_{B}F_{U}F_{B}F_{S}\bar{\bm{r}}_{0},
\end{equation}
For output mode $a$, the corresponding covariance matrix $\sigma_{a}$
and the mean vector $\bar{\bm{r}}_{a}$ are obtained by taking the
first two components of $\sigma_{5}$ and $\bar{\bm{r}}_{5}$,
\begin{align}
\sigma_{a} & =(\sigma_{5})_{12}=\left(\begin{array}{cc}
e^{-r}\Gamma_{+} & 0\\
0 & e^{r}\Gamma_{-}
\end{array}\right),\\
\bar{\bm{r}}_{a} & =(\bar{\bm{r}}_{5})_{12}=\sqrt{2}\alpha\cos\frac{\phi}{2}\!\left(\begin{array}{c}
1\\
0
\end{array}\right),
\end{align}
with $\Gamma_{\pm}\!=\!\cosh r\pm\sinh r\cos\phi$. Correspondingly,
for output mode $b$, 
\begin{align}
\sigma_{b} & =(\sigma_{5})_{34}=\left(\begin{array}{cc}
e^{3r}\Gamma_{+} & 0\\
0 & e^{-3r}\Gamma_{-}
\end{array}\right),\\
\bar{\bm{r}}_{b} & =(\bar{\bm{r}}_{5})_{34}=\sqrt{2}\alpha e^{-r}\sin\frac{\phi}{2}\!\left(\begin{array}{c}
0\\
1
\end{array}\right).
\end{align}
With these covariance matrices and mean vectors, we construct the
Wigner functions $W\!\left(x_{i},p_{i}\right)$ for the output modes
$i\!=\!a$, $b$. The intensity-difference signal is $\langle N_{-}\rangle\!=\!I_{a}-I_{b}$,
where 
\begin{equation}
I_{i}=\frac{1}{2}\!\iint\!\big(x_{i}^{2}+p_{i}^{2}\big)W(x_{i},p_{i})dx_{i}dp_{i},
\end{equation}
which describes the output intensity of mode $i$ up to a constant
factor $-1/2$. The intensity difference admits a natural interpretation
as the moment of inertia associated with the Wigner weight distribution.
As shown in Fig.~\eqref{fig:DS-MZI}, when $\alpha^{2}\!=\!\sinh^{2}\!r$,
the conventional scheme yields identical elliptical distributions
with equal output intensities $I_{a}\!=\!I_{b}$ and thus $\langle N_{-}\rangle\!=\!0$.
In contrast, the DS-MZI architecture produces a highly squeezed elliptical
distribution at the relevant output port, leading to an intensity
imbalance $I_{b}\!>\!I_{a}$ and $\langle N_{-}\rangle\!\neq\!0$.

\section*{Appendix D: Error propagation for imperfect photon-number-difference
detection \label{sec:AppendixD}}

  \makeatletter \renewcommand{\theequation}{D\arabic{equation}}\makeatother \setcounter{equation}{0} 

Here we derive the error-propagation formula for photon-number-difference
detection with non-unit efficiency $\eta$. The corresponding positive-operator-valued
measure (POVM) for photon counting is given by \citep{DallArno2010PRA,Lee2021CR}
\begin{equation}
\Pi_{\eta}(m)=\binom{a^{\dag}a}{m}\eta^{m}(1-\eta)^{a^{\dag}a-m},
\end{equation}
such that the detection probability is $p_{\eta}(m)\!=\!\langle\Pi_{\eta}(m)\rangle$,
where $\langle\bullet\rangle$ denotes the expectation value over
the detected states. Such non-ideal detection can be faithfully modeled
by an ideal detector preceded by a fictitious beam splitter with transmissivity
$\eta$, so that the noisy output operator is
\begin{equation}
\tilde{a}=\sqrt{\eta}a+\sqrt{1-\eta}v_{a},
\end{equation}
where $v_{a}$ denotes a vacuum environmental mode associated with
mode $a$. An analogous definition applies also to the output mode
$b$ under the assumption of identical detection efficiency $\eta$. 

The first and second moments of the measured photon number are given
by 
\begin{equation}
\langle\tilde{a}^{\dag}\tilde{a}\rangle=\sum_{m=0}^{\infty}m\langle\Pi_{\eta}(m)\rangle,
\end{equation}
and 
\begin{equation}
\big\langle(\tilde{a}^{\dag}\tilde{a})^{2}\big\rangle=\sum_{m=0}^{\infty}m^{2}\langle\Pi_{\eta}(m)\rangle.
\end{equation}
Using the operator identities \citep{DallArno2010PRA}
\begin{equation}
\sum_{m=0}^{\infty}\frac{m}{\eta}\Pi_{\eta}(m)=a^{\dag}a,
\end{equation}
and 
\begin{equation}
\sum_{m=0}^{\infty}\Big(\frac{m}{\eta}\Big)^{2}\Pi_{\eta}(m)=(a^{\dag}a)^{2}+\frac{1-\eta}{\eta}a^{\dag}a,
\end{equation}
we obtain 
\begin{equation}
\langle\tilde{a}^{\dag}\tilde{a}\rangle=\eta\langle a^{\dag}a\rangle,
\end{equation}
and 
\begin{equation}
\big\langle(\tilde{a}^{\dag}\tilde{a})^{2}\big\rangle=\eta^{2}\langle\Delta^{2}(a^{\dag}a)\rangle+\eta(1-\eta)\langle a^{\dag}a\rangle.
\end{equation}
Similar expressions hold for the output mode $b$.

Accordingly, the mean and variance of the measured photon-number difference
$\tilde{N}_{-}\!=\!\tilde{a}^{\dag}\tilde{a}-\tilde{b}^{\dag}\tilde{b}$
become
\begin{equation}
\big\langle\tilde{N}_{-}\big\rangle=\eta\langle N_{-}\rangle,
\end{equation}
and
\begin{equation}
\big\langle\Delta^{2}\tilde{N}_{-}\big\rangle=\eta^{2}\langle\Delta^{2}N_{-}\rangle+\eta(1-\eta)\langle N_{+}\rangle,
\end{equation}
where $N_{+}\!=\!a^{\dag}a+b^{\dag}b$ is the total output photon-number
operator. Here we have used the relation $\big\langle\tilde{a}^{\dag}\tilde{a}\tilde{b}^{\dag}\tilde{b}\big\rangle\!=\!\eta^{2}\big\langle a^{\dag}a\big\rangle\big\langle b^{\dag}b\big\rangle$,
which holds because the detection noises on the two output modes are
uncorrelated. Substituting these expressions into the standard error-propagation
formula 
\begin{equation}
\Delta\phi=\frac{\sqrt{\!\langle\Delta^{2}\tilde{N}_{-}\rangle}}{\big\vert\frac{d\langle\tilde{N}_{-}\rangle}{d\phi}\big\vert}
\end{equation}
directly yields the modified error-propagation formula Eq.~\eqref{eq:error}
used in the main text.
\end{document}